# Undulations in a weakly interacting mechanically generated molecular braid under tension

Dominic Lee[1,a.)]

[1]Department of Chemistry, Imperial College London, SW7 2AZ, London, UK

**Abstract**

We consider mechanically generated molecular braids composed of two molecules where long range interactions between them can be considered to be very weak. We describe a model that takes account of the thermal fluctuations of the braid, steric interactions between the molecules, and external mechanical forces. In this model, both sets of ends, of the two molecules, are considered to be separated by a fixed distance apart much larger than the radius of the braid. One set of ends are rotated to generate a braid of a certain number of pitches (or turns), while the other set remains fixed. This model may describe the situation in which the ends of each molecule are attached to a substrate and a magnetic bead; to the latter a pulling force and rotational torque can be applied. We discuss various aspects of our model. Most importantly, an expression for the free energy is given, from which equations, determining the various geometric parameters of the braid, can be obtained. By numerically solving these equations, we give predictions from the model for the external torque needed to produce a braid with a certain number of turns per bending persistence length, as well as the end to end extension of the two molecules for a given pulling force. Other geometric parameters, as well as the lateral force required to keep the ends of the two molecules apart, are also calculated.

## 1.Introduction

Molecular braids are present in nature; for instance, supercoiled DNA [1], braids formed of actin and tropomyosin within muscles [2], collagen fibres [3] and amyloids [4]; the latter being an important factor in neuro-degenerative illness. Recently, manmade molecular braids have been generated from metal organic compounds [5,6] which may have novel applications, whilst other molecules have been seen to self-assemble themselves into larger braided structures [7]. Therefore, the physics of molecular braiding is an intriguing subject for investigation.

Ground state elastic rod models, without intermolecular interactions were studied in [8,9,10,11]. The statistical mechanics of braiding of two molecular strands, including Debye screened electrostatic interactions, has been studied in [12,13,14], mainly in application to DNA. Most notably, Ref. [14] was the first to consider the statistical mechanics of the braiding of two molecules under tension. Recently the possibility of spontaneous braiding of

---

[a.)] Electronic Mail: domolee@hotmail.com

two helical molecules was investigated [15,16], caused by possible intermolecular interaction forces that depend on helix structure. These studies were primarily aimed at DNA, but such braiding has not yet been observed. Though, interestingly enough, the spontaneous braiding of three or more actin molecules has indeed been observed in the presence of high $Mg^{2+}$ concentrations [17], which suggests, that for these molecules, forces that depend on the helical structure of the molecules may matter under certain conditions. The work of Refs. [15] and [16] could be indeed be adapted to study the formation of such actin braids.

Molecular braiding has also been studied on the experimental side. Experiments involving the mechanical braiding of two molecules of DNA have been performed, in which the two molecules are attached to a magnetic bead and a substrate [18,19]. Both the pulling force and the number of turns of the bead are controlled by varying a magnetic field. However, this type of experiments should not be restricted to just DNA alone. These experiments allow for the opportunity to understand the interplay between interactions, fluctuations and external forces in braided structures of different types of semi-flexible rod like molecules. In this paper we focus on a model describing such mechanical braiding between molecules, first considered in Ref. [20]. However, here we consider any non-steric interactions between the two molecules sufficiently weak to be neglected. This model uses analytical results from Ref. [21], which were also used to include the effect of undulations in the mechanical braiding of DNA molecules [20], to describe the braiding experiments of Ref. [19]. An attractive feature of this model, without intermolecular interactions other than steric ones, is that through rescaling there are universal features, which may describe molecules of any bending persistence length.

The work is structured as follows. In the subsequent section, we describe the model, as well as geometrical and topological considerations that we need to take into account when describing the braided section of the two molecules. We give an expression for the free energy of the braid, derived in Ref. [21], as well as demonstrating how, from this expression, equations determining the braid radius, the tilt angle (the angle between the two tangent vectors of the molecules in the braid), end to end distance and number of braid turns can all be derived. In our treatment, the bending energy is averaged over thermal fluctuations in the radius of the braid. We discuss a simplifying approximation where bending energy is not averaged; rather, the bending energy of the average braid structure is used, as was considered in previous studies. We also discuss possible coexistence between braided sections and 'bubbles', where the two molecules lie far apart in an unbraided state, first considered in Ref. [14]. In the Results section, we look at two choices of separation length between the ends of the two molecules forming the braided structure. In the first set of results, we present values of the moment that needs to be applied to produce a certain number of braid turns, for various values of the pulling force and the ratio of hard core radius of the two molecules to their bending persistence length. Next, we plot the end to end extension of the two molecules against the number of braid turns. We show how geometric characteristics of the braid vary with the number of braid pitches: the contour length braided section, the braid radius and the tilt angle. Finally, we show the lateral force required to keep the molecules apart, which might conceivably be measured using state of the art micro manipulation experiments of form considered in Ref. [22]. We see that the simple approximation agrees well with the full model when the pulling force is sufficiently high or the number braid turns is large enough, but it does not agree well when considering the coexistence region. In discussion section we discuss our findings and point to future work.

## 2. Theory

### 2.1 General Considerations

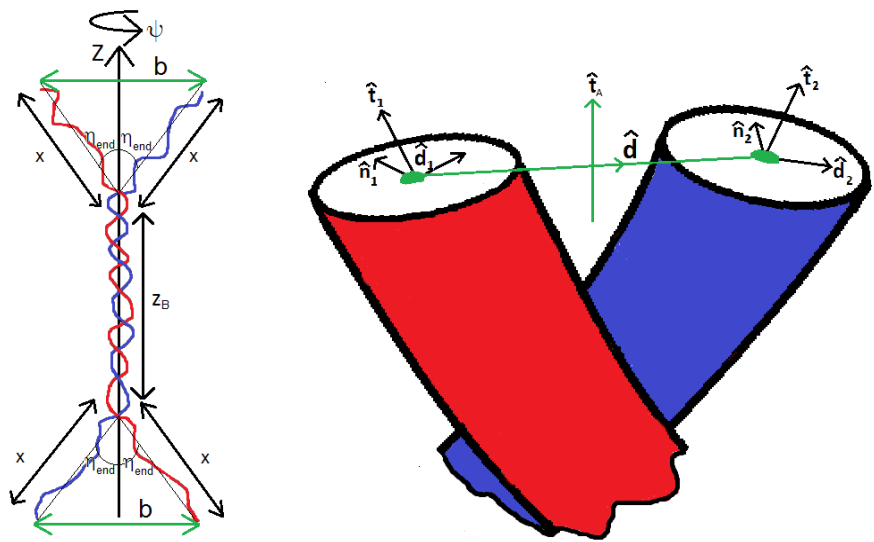

Fig.1. Schematic pictures of the braided configuration of the two molecules. The left hand picture shows the global configuration of two molecules. The red (lighter) line denotes one molecule, while the blue (darker) line denotes the other one. Both sets of molecular ends are separated distance $b$ apart and one set of ends are rotated an angle $\psi$ with respect to the other set. To generate this configuration, two of the ends may be attached to a magnetic bead and the other two ends to a substrate. In the second picture we show the tangent vectors ($\hat{\mathbf{t}}_1(\tau)$ and $\hat{\mathbf{t}}_2(\tau)$) of the two molecular centre lines, the vector $\hat{\mathbf{d}}(\tau)$ that lies along a line connecting the two molecular centre lines (shown in green) and the tangent vector of the braid centre line that define the local configuration of the braid. Also shown are the vectors $\hat{\mathbf{n}}_\mu(\tau)$ and $\hat{\mathbf{d}}_\mu(\tau)$, which are defined in [16] and [21], that define the braid frames [23] of the two molecules. When $R'(\tau) = 0$ we have that $\hat{\mathbf{d}}_1(\tau) = \hat{\mathbf{d}}_2(\tau) = \hat{\mathbf{d}}(\tau)$.

We consider two DNA molecules of the same length $L$, which are mechanically braided about each other. We label the molecules with index $\mu = 1, 2$, and we label the ends of each molecule $\mu$ with index $i = 1, 2$, so that each end has the label $\{\mu, i\}$. Ends $\{1,1\}$ and $\{2,1\}$ are held a distance $b$ apart, as well the ends $\{1,2\}$ and $\{2,2\}$. We place our z-axis though the mid points of two connecting lines between the two sets of ends. These lines connecting $\{1,1\}$ with $\{2,1\}$ and $\{1,2\}$ with $\{2,2\}$ are chosen to be perpendicular to the z-axis, so that the two DNA are constrained within a rectangle like configuration (see Fig. 1). We allow for the ends $\{1,2\}$ and $\{2,2\}$ to rotate around the z axis an angle $\psi$, away from the configuration where the two molecular centre lines are parallel. Here, the ends $\{1,1\}$ and $\{2,1\}$ are considered fixed. The number of rotations of ends $\{1,2\}$ and $\{2,2\}$ about the z-axis is given by $n = \psi / 2\pi$. A pulling force $F = k_B T F_R / l_b$ (where $l_b$ is the bending persistence length of the two molecules, defined below) is applied on each molecule along the z-axis and a moment $M = k_B T M_R$ is applied about the z-axis to rotate the molecular ends, producing a central braided section with $N$ braid pitches. Both $n$ and $N$ are related to each other through

$$n \approx N + \frac{\text{sgn}(n)}{2}, \qquad (1.1)$$

where we suppose that the braid diameter $R$ is such that $R \ll b$. Here, positive values of $N$ and $n$ correspond to left handed braiding, whereas negative values correspond to right handed braiding. The total vertical distance between the two ends we define as $z_T$, which is a function of both the force and moment. The contour length of both molecules contributing to the braided section is given by $L_b$. The unbraided parts (end pieces) of each molecule are assumed to have equal contour length $(L - L_b)/2$. To describe the bending fluctuations of the centre lines of these unbraided parts we use worm like chains, and the average positions of these parts of the molecular centre lines are described by straight lines of length $x$. These lines connect the molecular ends to, and the tangent vectors of those lines match with, the ends of the braided section. This allows us to write the end to end distance

$$z_T \approx z_B + \sqrt{(2x)^2 - b^2}. \qquad (1.2)$$

We may then write down the following free energy to describe our system

$$\mathcal{F}_T = 2(L - L_b) f_{wlc} + \mathcal{F}_{Braid}, \qquad (1.3)$$

where $f_{wlc}$ is a worm like chain free energy density, for fixed $F$ and $b$, and $\mathcal{F}_{Braid}$ is the free energy of the braided section. We will start by discussing the form of $f_{wlc}$ and then we will discuss the braided part of the construction.

### 2.2 The end pieces

Here, we obtain an expression for $f_{wlc}$ and relate $L - L_b$ to $x$. The lateral force holding the two sets of ends distance $b$ apart is given by

$$F_b = \tan\left(\frac{\eta_{end}}{2}\right) F_R, \tag{1.4}$$

and the total force acting along the tangents of the average positions of the two centre lines is

$$F_T = \frac{1}{2}\sqrt{F_R^2 + F_b^2} = \frac{F_R}{2\cos\left(\frac{\eta_{end}}{2}\right)}, \tag{1.5}$$

where $\eta_{end}$ is the angle between the two tangent vectors of the molecular centre lines at the end of the braid section (see below).

For large forces, we can write down the worm like chain free energy, $g_{wlc}(F_T)$ for a fixed $F_T$ ensemble [24]

$$\frac{g_{wlc}(F_T) l_p}{k_B T} \approx -F_T + (F_T)^{1/2}. \tag{1.6}$$

Also, the average end to end distance $x$ of an end section is related to its contour length of $L - L_b$ through the expression [24]

$$\frac{2x}{L - L_b} \approx 1 - \frac{1}{2(F_T)^{1/2}}. \tag{1.7}$$

Combining Eqs. (1.5) and (1.7) allows us to write for $L_b$

$$L_b \approx L - \left(1 + \left(\frac{\cos\left(\frac{\eta_{end}}{2}\right)}{2F_R}\right)^{1/2}\right) \frac{b}{\left|\sin\left(\frac{\eta_{end}}{2}\right)\right|}, \text{ when } \left(1 + \left(\frac{\cos\left(\frac{\eta_{end}}{2}\right)}{2F_R}\right)^{1/2}\right) \frac{b}{\left|\sin\left(\frac{\eta_{end}}{2}\right)\right|} < L, \quad (1.8)$$

otherwise $L_b = 0$, and we just have end pieces. When $L_b = 0$ we have $N = 0$ and $n = \pm 1/2$. We do not consider the region where $-1/2 < n < 1/2$.

The free energy for a fixed $F_R$ and $b$ ensemble is obtained from Eq. (1.7) through a Legendre transformation

$$\frac{f_{wlc}(b,F)}{k_B T}(L - L_b) \approx \frac{g_{wlc}(F_T)}{k_B T}(L - L_b) + \frac{F_b b}{2l_p}. \quad (1.9)$$

Then, using Eqs.(1.4), (1.5), (1.6) and (1.9) we can write

$$\frac{2 f_{wlc}(b,F)(L - L_b)}{k_B T} \approx -\frac{F_R b}{l_p \left|\sin\left(\frac{\eta_{end}}{2}\right)\right|} \left(\cos\left(\frac{\eta_{end}}{2}\right) - \left(\frac{1}{2F_R \cos\left(\frac{\eta_{end}}{2}\right)}\right)^{1/2}\right). \quad (1.10)$$

The first term in Eq. (1.10) represents a mechanical work term in terms of $F_R$, in absence of thermal fluctuations. This is the work done stretching out straight end pieces, at fixed $b$. The second term represents the worm like chain correction to it due to thermal fluctuations.

## 2.3 Braid Geometry

Now, we describe the geometry of the braid. We start by defining a line of length $R(\tau)$ that connects the centrelines of the two molecules, where $\tau$ runs from $-L_A/2$ to $L_A/2$; $L_A$ is the total contour length of the braid axis (defined below), which is indeed a function of $L_b$, the average contour length of the molecular centre lines within the braided part. We define a unit vector $\hat{\mathbf{d}}(\tau)$ that points along this line connecting the two molecular centre lines. The midpoint of the connecting line is bisected by the braid axis, which is perpendicular to it, with a tangent vector (see Fig. 1)

$$\hat{\mathbf{t}}_A(\tau) = \frac{dx_A(\tau)}{d\tau}\hat{\mathbf{i}} + \frac{dy_A(\tau)}{d\tau}\hat{\mathbf{j}} + \sqrt{1 - \left(\frac{dx_A(\tau)}{d\tau}\right)^2 - \left(\frac{dy_A(\tau)}{d\tau}\right)^2}\hat{\mathbf{k}}. \quad (1.11)$$

Here $x_A(\tau)$ and $y_A(\tau)$ are lateral displacements of the braid axis away from a straight line configuration of the braid axis coincident with the z-axis. The position vector of the braid axis is given through

$$\mathbf{r}_A(\tau) = x_A(\tau)\hat{\mathbf{i}} + y_A(\tau)\hat{\mathbf{j}} + \int_{-L_A/2}^{\tau} d\tau' \sqrt{1 - \left(\frac{dx_A(\tau')}{d\tau'}\right)^2 - \left(\frac{dy_A(\tau')}{d\tau'}\right)^2} \hat{\mathbf{k}}. \qquad (1.12)$$

The requirements that $\hat{\mathbf{d}}(\tau).\hat{\mathbf{t}}_A(\tau) = 0$ and $|\hat{\mathbf{d}}(\tau)| = 1$, allow us to derive an exact parameterization of $\hat{\mathbf{d}}(\tau)$, which is given in Ref. [21]. However, for small values of $x'_A(\tau)$ and $y'_A(\tau)$ (here the prime refers to differentiation with respect to argument) we may write (see Ref. [21])

$$\hat{\mathbf{d}}_A(\tau) \approx \cos\theta(\tau)\hat{\mathbf{i}} + \sin\theta(\tau)\hat{\mathbf{j}}, \qquad (1.13)$$

where $\theta(\tau)$ is an angle that describes the rotation of the molecular centre lines, forming the braid, around the $z$-axis. We can then define the position vectors along the centre lines of the two molecules in the braid as

$$\mathbf{r}_1(\tau) = \mathbf{r}_A(\tau) - \frac{R(\tau)\hat{\mathbf{d}}_A(\tau)}{2}, \qquad \mathbf{r}_2(\tau) = \mathbf{r}_A(\tau) + \frac{R(\tau)\hat{\mathbf{d}}_A(\tau)}{2}. \qquad (1.14)$$

We can also express the trajectories in Eq. (1.14) as $\mathbf{r}_1(\tau) = \tilde{\mathbf{r}}_1(s_1)$ and $\mathbf{r}_2(\tau) = \tilde{\mathbf{r}}_1(s_2)$. The coordinate $s_\mu$ is the unit arc length coordinate of the molecular centre lines of molecule $\mu$, which runs from $-L_\mu/2$ to $L_\mu/2$, where $L_\mu$ is the contour length of molecule $\mu$ forming the braided section. On averaging over fluctuations of the braid axis we have that $\langle L_\mu \rangle = L_b$, and we assume that the molecules are long enough that fluctuations in $L_\mu$ about $L_b$ can be considered as very small. From the definition of unit arc length we have that

$$\hat{\mathbf{t}}_\mu(s_\mu) = \frac{d\tilde{\mathbf{r}}_\mu(s_\mu)}{ds_\mu} \quad \text{and} \quad |\hat{\mathbf{t}}_\mu(s_\mu)| = 1. \qquad (1.15)$$

Expressions for $\hat{\mathbf{t}}_1(s_\mu)$ and $\hat{\mathbf{t}}_2(s_\mu)$ can be obtained through Eqs. (1.11)-(1.15), and are presented in Ref. [21]. One last important geometric parameter is the tilt angle $\eta(\tau)$ of the braid. This is defined through the dot product of the tangent vectors of the two molecular centre lines, which reads as

$$\hat{\mathbf{t}}_1(s_1(\tau)).\hat{\mathbf{t}}_2(s_2(\tau)) = \cos\eta(\tau), \qquad (1.16)$$

also note that $\eta(L_b/2) = \eta(-L_b/2) = \eta_{end}$.

## 2.4 Braid Linking Number, Twist and Writhe

The fixed number of turns of the braid ends, $N$ is a conserved topological quantity. This is equivalent to $-Lk_b$, where $Lk_b$ is what we term as the braid linking number [20,21], which is sometimes referred to as the catenation number. This quantity (in an analogous way to single molecules) can be related to two other quantities, which we call Braid Twist and Braid Writhe. These are related to the braid linking number through the Fuller-White formula

$$-N = Lk_b = Tw_b + Wr_b. \tag{1.17}$$

The braid twist is approximately given by the expression [21]

$$Tw_b \approx \frac{1}{2\pi} \int_{-L_A/2}^{L_A/2} d\tau \frac{d\theta(\tau)}{d\tau}, \tag{1.18}$$

and the Braid Writhe is evaluated by the following Gauss's Integral

$$Wr_b = \frac{1}{4\pi} \int_{-L_A/2}^{L_A/2} d\tau \int_{-L_A/2}^{L_A/2} d\tau' \frac{(\mathbf{r}_A(\tau) - \mathbf{r}_A(\tau')) \cdot \hat{\mathbf{t}}_A(\tau) \times \hat{\mathbf{t}}_A(\tau')}{|\mathbf{r}_A(\tau) - \mathbf{r}_A(\tau')|^3}. \tag{1.19}$$

For a further discussion of these two quantities, see Ref. [21]. In what follows the topological constraint that fixes $Lk_b$ will be imposed through a Lagrange multiplier so that its average value is constrained (see next subsection).

## 2.5 Energy functional of the braid

The energy functional that we consider in the partition function consists of three contributions. The first of these is an elastic energy contribution of the form

$$\frac{E_{els}}{k_B T} = \frac{l_p}{2} \int_{-L_1/2}^{L_1/2} ds_1 \left(\frac{d\hat{\mathbf{t}}_1(s_1)}{ds_1}\right)^2 + \frac{l_p}{2} \int_{-L_2/2}^{L_2/2} ds_2 \left(\frac{d\hat{\mathbf{t}}_2(s_2)}{ds_2}\right)^2, \tag{1.20}$$

where $s_1$ and $s_2$ are the unit arc length coordinates of molecules 1 and 2 of lengths $L_1$ and $L_2$ respectively making up the braid. The bending persistence length $l_p$ is defined as $l_p = B/k_B T$ where $B$ is the bending rigidity of the two molecules, which we have supposed to be identical for both.

The second contribution is from a work term that contains an external moment and pulling force acting on the braid (Lagrange multipliers) that constrain the average braid linking number and the end to end distance $z_B$ of the braid. This contribution to the energy functional is written as

$$E_W = -Fz_B - 2\pi ML k_b. \tag{1.21}$$

Both the average values of $z_B$ and $N$ are found through derivatives of the free energy of the braid $\mathcal{F}_{Braid}$ with respect to $F$ and $M$, namely

$$z_B \approx \langle z_B \rangle = -\frac{\partial \mathcal{F}_{Braid}}{\partial F} \quad \text{and} \quad N \approx \langle N \rangle = \frac{1}{2\pi}\frac{\partial \mathcal{F}_{Braid}}{\partial M}. \tag{1.22}$$

Lastly, there is a contribution from steric interactions of the two molecules. We will suppose that both molecules can be approximated as smooth cylinders, and that both molecules have the same hard core radius $a$. To model the effect of steric interactions we introduce a harmonic confining potential of the form

$$E_H = \frac{\alpha_H}{2}\int_{-L_A/2}^{L_A/2} d\tau \left(R(\tau) - R_0\right)^2 = \frac{\alpha_H}{2}\int_{-L_A/2}^{L_A/2} d\tau \delta R(\tau)^2, \tag{1.23}$$

as in the manner of Helfich [25], and used in Ref. [26] for DNA assemblies. Here $R_0$ is the average value of $R(\tau)$, i.e. $R_0 = \langle R(\tau) \rangle$. Also, in the elastic energy we introduce the cut-offs $d_{max}$ and $d_{min}$. These represent the maximum and minimum values that $\delta R(\tau)$ is limited to by steric interactions. The parameter $\alpha_H$ is also dependent on these two values; we will later show roughly what form it should take to model the steric interactions. When $\delta R(\tau) > d_{max}$ and $\delta R(\tau) < d_{min}$, we replace $\delta R(\tau)$ with $d_{max}$ and $d_{min}$, respectively, in explicit expressions for $E_{els}$ and $E_W$ in terms of $R$, $\eta$, $x_A$ and $y_A$, derived from Eqs. (1.20) and (1.21) (see Ref. [21]). We leave the derivatives of $\delta R(\tau)$ with respect to arc-length coordinate untouched. These cut-offs prevent unphysical values in $E_{els}$ and $E_W$ arising from two centre lines coming too close, or too far apart, than what steric interactions would allow, when we use Eq. (1.23) to model them. This procedure was first used for assemblies of rod like molecules [26,27]. For braids, this procedure is explicitly demonstrated in Ref. [21], and an example of it can also be found in Ref. [16] for spontaneous braiding.

For tightly wound braids confined by strong attractive interactions, we have argued that $d_{max}$ should be larger than $R_0 - 2a$ [16]. However, in this case, we expect a greater degree of fluctuation in the braid pitch. Also, we have only steric contributions contributing to $\langle \delta R(\tau)^2 \rangle$ and $R_0 - 2a$ is much larger than for the braiding considered in Ref. [16]. Therefore, we think that the better choice here may be $d_{max} \approx d_{min} = R_0 - 2a$, which is what we indeed make. This choice was also used in Refs. [12] and [14]. Subsequently, to model the steric confinement, we also choose $\alpha_H$ such that

$$\langle \delta R(\tau)^2 \rangle \approx R_0 - 2a, \tag{1.24}$$

yielding the following expression for $\alpha_H$

$$\alpha_H = \frac{1}{2^{5/3}(R_0 - 2a)^{8/3} l_p^{1/3}}. \tag{1.25}$$

## 2.6 Free energy expressions, average number of braid turns and braid extension

From the three contributions to the energy functional discussed in the previous subsection, we find the following expression for the free energy (for details of the calculation and the approximations used see Ref. [21])

$$\frac{\mathcal{F}_{Braid}}{k_B T L_b} = \frac{\mathcal{E}_b}{l_p} \approx \frac{1}{l_p} \left[ \left( \frac{F_R}{2\cos(\eta_0/2)} \right)^{1/2} + \frac{1}{2^{2/3}(\tilde{R}_0 - 2\tilde{a})^{2/3}} + \frac{\tilde{\alpha}_\eta^{1/2}}{2^{1/2}} \right.$$

$$-2^{1/3} \left( \frac{f_1(\tilde{R}_0;\tilde{a})}{\tilde{R}_0^2} \sin^2\left(\frac{\eta_0}{2}\right) + \frac{M_R f_2(\tilde{R}_0;\tilde{a})}{4\tilde{R}_0} \sin\left(\frac{\eta_0}{2}\right) \right)^{-1} \left( \frac{\tilde{R}_0 - 2\tilde{a}}{2} \right)^{2/3} - \frac{1}{16} \frac{M_R^2}{(2F_R)^{1/2} \cos\left(\frac{\eta_0}{2}\right)^{7/2}}$$

$$\left. +4\sin^4\left(\frac{\eta_0}{2}\right) \frac{f_1(\tilde{R}_0;\tilde{a})}{\tilde{R}_0^2} + \frac{2M_R f_2(\tilde{R}_0;\tilde{a})}{\tilde{R}_0} \sin\left(\frac{\eta_0}{2}\right) - F_R \cos\left(\frac{\eta_0}{2}\right) \right], \tag{1.26}$$

where $\eta_0 = \langle \eta(\tau) \rangle$ is the mean value of the tilt angle defined through Eq. (1.16). Here, we have introduced the rescaling, $\tilde{R}_0 = R_0 / l_p$, $\tilde{a} = a / l_p$ and we have that

$$\tilde{\alpha}_\eta = \left[ \frac{4 f_1(\tilde{R}_0;\tilde{a})}{\tilde{R}_0^2} \left( 3\cos^2\left(\frac{\eta_0}{2}\right) \sin^2\left(\frac{\eta_0}{2}\right) - \sin^4\left(\frac{\eta_0}{2}\right) \right) + \frac{F_R}{4} \cos\left(\frac{\eta_0}{2}\right) - \frac{M_R f_2(\tilde{R}_0;\tilde{a})}{2\tilde{R}_0} \sin\left(\frac{\eta_0}{2}\right) \right], \tag{1.27}$$

as well as

$$f_1(\tilde{R}_0;\tilde{a}) = \frac{\tilde{R}_0^2}{(\tilde{R}_0-2\tilde{a})\sqrt{2\pi}}\int_{2\tilde{a}-\tilde{R}_0}^{\tilde{R}_0-2\tilde{a}}\frac{dx}{(\tilde{R}_0+x)^2}\exp\left(-\frac{x^2}{2(\tilde{R}_0-2\tilde{a})^2}\right) + \frac{1}{2}\left(\frac{\tilde{R}_0^2}{4\tilde{a}^2}+\frac{\tilde{R}_0^2}{4(\tilde{R}_0-\tilde{a})^2}\right)\left(1-\mathrm{erf}\left(\frac{1}{\sqrt{2}}\right)\right),$$
(1.28)

$$f_2(\tilde{R}_0;\tilde{a}) = \frac{\tilde{R}_0}{(\tilde{R}_0-2\tilde{a})\sqrt{2\pi}}\int_{2\tilde{a}-\tilde{R}_0}^{\tilde{R}_0-2\tilde{a}}\frac{dx}{(\tilde{R}_0+x)^2}\exp\left(-\frac{x^2}{2(\tilde{R}_0-2\tilde{a})^2}\right) + \frac{1}{2}\left(\frac{\tilde{R}_0}{2\tilde{a}}+\frac{\tilde{R}_0}{2(\tilde{R}_0-\tilde{a})}\right)\left(1-\mathrm{erf}\left(\frac{1}{\sqrt{2}}\right)\right).$$
(1.29)

An important assumption in the derivation of Eq. (1.26) that $R'(\tau)$, $\eta'(\tau)$, $x'_A(\tau)$, $y'_A(\tau)$, and $\eta(\tau)-\eta_0$ are all small. For these assumptions to hold we require that $\langle R'(\tau)^2\rangle = (\tilde{R}_0-2\tilde{a})/2^{1/3} \ll 1$, $l_p \gg h$, $\tilde{F}_R \gg 1$ and $\tilde{\alpha}_\eta \gg 1$, where $h$ is a length scale below which the continuum description of bending elasticity, Eq. (1.20), can be considered no longer valid.

Next, using Eqs. (1.3), (1.10) and (1.26), we can also write down an expression for the total free energy of the two molecules

$$\frac{\mathcal{F}_T l_p}{k_B T L} \approx \mathcal{E}_b - \frac{\beta}{\left|\sin\left(\frac{\eta_{end}}{2}\right)\right|}\left(F_R\cos\left(\frac{\eta_{end}}{2}\right) + \mathcal{E}_b - \left(\frac{F_R}{2\cos\left(\frac{\eta_{end}}{2}\right)}\right)^{1/2} + \mathcal{E}_b\left(\frac{\cos\left(\frac{\eta_{end}}{2}\right)}{2F_R}\right)^{1/2}\right),\quad (1.30)$$

where $\beta = b/L$. We find for $\tilde{F}_R \gg 1$ that the value of $\eta_{end}$ that minimizes Eq. (1.30) is

$$\cos\left(\frac{\eta_{end}}{2}\right) \approx -\frac{F_R}{\mathcal{E}_b} - \left(\frac{2}{F_R}\right)^{1/2}\left(-\frac{F_R}{\mathcal{E}_b}\right)^{3/2}.\quad (1.31)$$

The equations that determine the geometric parameters $\eta_0$ and $\tilde{R}_0$ are then got through the minimization conditions

$$\frac{\partial \mathcal{E}_b}{\partial \tilde{R}_0} = 0,\qquad \frac{\partial \mathcal{E}_b}{\partial \eta_0} = 0.\quad (1.32)$$

The explict equations that both $\tilde{R}_0$ and $\eta_0$ satisfy, obtained from Eqs. (1.26) and (1.32), can found in Section 9 of Ref. [21].

From Eqs. (1.22) and (1.26), $\tilde{N}$ the number of braid turns (or braid linking number) contained in each persistence length along the braided section is obtained through

$$\tilde{N} = \frac{Nl_p}{L} = \frac{\gamma}{2\pi}\frac{\partial \mathcal{E}_b}{\partial M}\quad (1.33)$$

where $\gamma = L_b / L$. The end to end distance divided by the contour lengths of the two molecules $\tilde{z}_T$ (from Eqs (1.22) and (1.26)) is given by the expression

$$\tilde{z}_T = \frac{z_T}{L} \approx \tilde{z}_B + \sqrt{(1-\gamma)^2 \left(1 - \left(\frac{\cos(\eta_{end}/2)}{2\tilde{F}_R}\right)^{1/2}\right)^2 - \beta^2}, \quad (1.34)$$

where

$$\tilde{z}_B = \frac{z_B}{L} = -\gamma \frac{\partial \mathcal{E}_b}{\partial F_R}. \quad (1.35)$$

Explicit expressions for $\tilde{N}$ and $\tilde{z}_B$ are readily derived from Eq. (1.26), (1.33) and (1.35) are also given in Ref. [21]. As well as using the full free energy (Eq. (1.26)) and the resulting equations for geometric parameters by differentiating it, we may also make a simplifying approximation. This approximation supposes that $f_1(\tilde{R}_0) \approx 1$ and $f_2(\tilde{R}_0) \approx 1$. Thus, it is simply got by replacing $f_1(\tilde{R}_0)$ and $f_2(\tilde{R}_0)$ with 1 in Eqs. (1.26) and (1.27). Then, resulting expressions for $\eta_0$, $R_0$, $\tilde{z}_B$ and $\tilde{N}$ are derived through Eqs. (1.32), (1.33) and (1.35). This approximation was used in Ref. [20], when interaction forces between molecules were supposed to be weak. We will test how well this approximation works against the full expressions in the results section.

## *2.7 Coexistence of braided sections with unbraided sections*

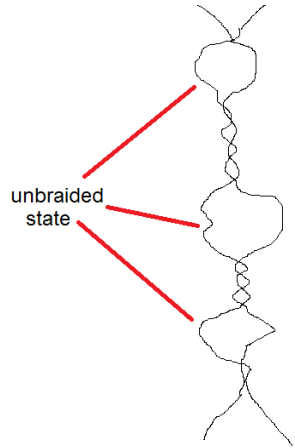

Fig 2. Schematic picture of the two molecules in the coexistence region. Here, braided sections coexist with 'bubbles'; sections where the two molecules are apart from each other and unbraided, highlighted in the picture. As $N$ gets smaller, more of the molecules become unbraided in such bubbles. When a critical number of turn is reached $N = N_c$ the 'bubbles' disappear.

Here, we consider that there could be regions, or bubbles, where the molecules are unbraided and far apart, as shown schematically in Fig. 2. Coexistence between braided and unbraided states was originally discussed in Ref. [14]. In these regions we assume that the two molecules behave as two worm like chains, far apart from each other, with pulling force $F/2$ acting on each of them. Then, for this unbraided state, the free energy density of the two worm like chains simply reads as

$$\frac{l_p g_{unbraid}(F_R)}{k_B T} = -F_R + \sqrt{2F_R}. \tag{1.36}$$

If $-\mathcal{E}_b$ is sufficiently large we have no 'bubbles', a purely braided state; but when we have that

$$\frac{l_p g_{unbraid}(F_R)}{k_B T} = \mathcal{E}_b, \tag{1.37}$$

we should have coexistence between the two states, as shown in Fig. 2. This coexistence happens at a critical value of the applied moment $M_R = M_c$, which is the solution to Eq. (1.37) along with Eqs. (1.26) and (1.32). At one end of the coexistence region we have a fully braided state (no 'bubbles'); the number of braid turns per persistence length $\tilde{N}_c$ is simply got from Eq (1.33) with $M_R = M_c$. At the other end $N = 0$; the braided sections are almost gone and bubbles predominate, when the molecules are considered to be infinitely long. However, for real molecules of finite length, particular care should be taken with the predictions of the model for small values of $N$. For molecules of finite length, to capture the small $N$ behaviour, near the end of the coexistence line, will require a more detailed theory taking account: i.) the bending energy required to create 'domain walls' between two states and a more accurate description of the bubbles, ii.) better treatment of the end sections in the model, iii.) the conservation of molecular contour length shared between braided and unbraided sections.

In our simple model the coexistence region both $\eta_{end}$ and the total contour length of braided and unbraided sections, $L_b$ remains constant. The rescaled extension of the braided section is now written for $0 < N \leq N_c$ as

$$\tilde{z}_B(N) = -\gamma(N_c) \left( \left. \frac{\partial \mathcal{E}_b}{\partial F_R} \right|_{N=N_c} \frac{N}{N_c} + \frac{N_c - N}{N_c} \frac{l_p}{k_B T} \frac{\partial g_{unbraid}(F_R)}{\partial F_R} \right)$$

$$= \gamma(N_c) \left( \tilde{z}_B(N_c) \frac{N}{N_c} + \frac{N_c - N}{N_c} \left( 1 - \frac{1}{\sqrt{2F_R}} \right) \right). \tag{1.38}$$

## 3. Results

### *3.1 Applied Moment*

We start by looking at the applied moment, $M$ as a function of the number of braid turns per persistence length. In Fig. 3 we show how this relationship changes as we vary the pulling force. Since, we consider weakly interacting braids with no chiral terms that prefer a particular braided handedness, all the curves have a $\tilde{N} \to -\tilde{N}$ and $M_R \to -M_R$ symmetry. There is also discontinuity at $\tilde{N}=0$ in the curves. This stems from the fact that the braid must have a smaller value of $z_T$ than the two molecules at $n=0$. Therefore, to move from the latter parallel configuration into the formation of the braided state, at $\tilde{N}=0$, requires mechanical work against the pulling force. This necessary work has to be provided by a finite value of the moment at $\tilde{N}=0$, the sign of which depends on the handedness of the braid. Indeed, as the pulling force values are increased, the size of discontinuity increases. This is due the increase in the amount of work that needs to be done against the pulling force to start to build a braid.

In Fig. 3a the value $\beta=b/L=0.1$ is used. Here, at a low value of $\tilde{N}$ (the number of braid turns in one persistence length), we see coexistence between the braided and unbraided states (see Subsection 2.7). In the coexistence region, as discussed previously, the value of $M_R$ stays constant at a critical value $M_c$. Coexistence does not occur at larger values of $\beta$, as seen in Fig. 3b for $\beta=0.35$. For it to occur there must be a stable braided configuration formed at $M_R = M_c$. The value of $M_c$ does not depend on $\beta$; however the contour length of the braided section $L_b$ (for the purely braided state) does. The critical energy density $\mathcal{E}_b(M_c)$ yields a value of $\eta_{end}(M_c)$ determined by Eq. (1.31). By setting $L_b=0$ in Eq. (1.8), we see that there must be a minimum value of $\eta_{end}$ for there to be a braided section, $\eta_{min}(\beta)$. By increasing $\beta$, one increases $\eta_{min}$ value until one reaches a critical value $\beta = \beta_c$, where $\eta_{min}(\beta_c) = \eta(M_c)$, and coexistence is lost. In other words, as we move from $\beta=0$ to $\beta=\beta_c$ the contour length $L_b$ for the braided configuration at the endpoint of the coexistence line becomes smaller until $L_b=0$ is reached and the braided state is eventually destabilized at $M_c$, and coexistence is lost.

In Fig. 3a we also observe that as we increase the force the coexistence regions also go away. At low forces, coexistence with the unbraided state helps to increase the entropy of the system, at a fixed value of $\tilde{N}$. When we increase the pulling force we reduce the amount of entropy that sections of the two molecules can gain by being in the unbraided state; coexistence with the unbraided state reduces the free energy less from the value of a purely braided state. Also, increasing the pulling force decreases the value of $\eta_{end}$ (see Eq. (1.8)) and so $L_b$; $\eta_{end}(M_c)$ decreases as the pulling force increases. At the point where the pulling force is sufficiently high ($F = F_c$) that $\eta_{min}(F_c) = \eta(M_c)$ the coexistence region again vanishes.

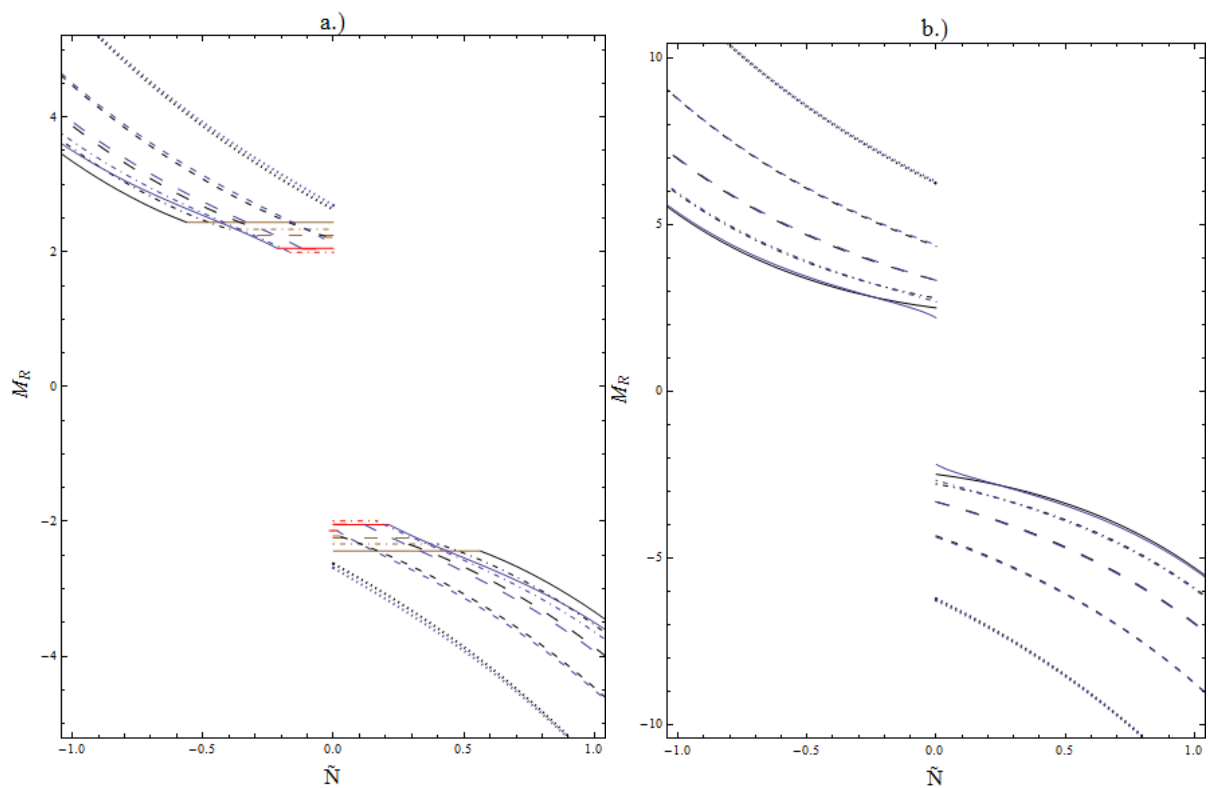

Fig.3. The moment $M$ (in units of $k_B T$) that needs to be applied to produce a certain number of braid turns per persistence length $\tilde{N}$. In these plots, we choose the value of $\tilde{a} = a/l_p = 1/50$. In a.) we choose $\beta = b/L = 0.1$ and in b.) $\beta = b/L = 0.35$. The black and brown (darker) curves correspond to the full model, whereas the red and blue (lighter) curves correspond to the simplifying approximation discussed at the end of Subsection 2.8. The red and brown curves correspond to coexistence region between the braided state and upbraided state, where the molecules are far apart, at a value $|M| = |M|_c$. The solid, dot-dashed, long dashed, medium dashed and dotted correspond to rescaled pulling force values $\tilde{F}_R = 25, 50, 100, 200, 400$, respectively. We see that the simple approximation works well at sufficiently large pulling force and braid turns. It does not agree well with the full model for the coexistence regions.

In both plots we compare the curves that are generated using the full expression for the braid free energy (Eq. (1.26)) with those of the simplifying approximation (see the end of Subsection 2.6). We see that the approximation doesn't agree with the full model at low forces and moments; it seems to considerably underestimate the size of the coexistence region. However, at large enough forces, and always for $\beta = 0.35$, agreement between the two is excellent. Indeed, the important factor that determines agreement between the two is the size of $R_0 - 2a$. If it is too large, then clearly, the simple approximation will not work

(see Eqs. (1.28) and (1.29)). In cases where the agreement is not good, the full model should provide a more accurate description. This is because if $R(\tau)$, the distance between the two centre lines, is fluctuating considerably, we would expect that an average value of the bending energy is better estimate of the bending contribution to the free energy than an unaveraged value fixed at $R_0$.

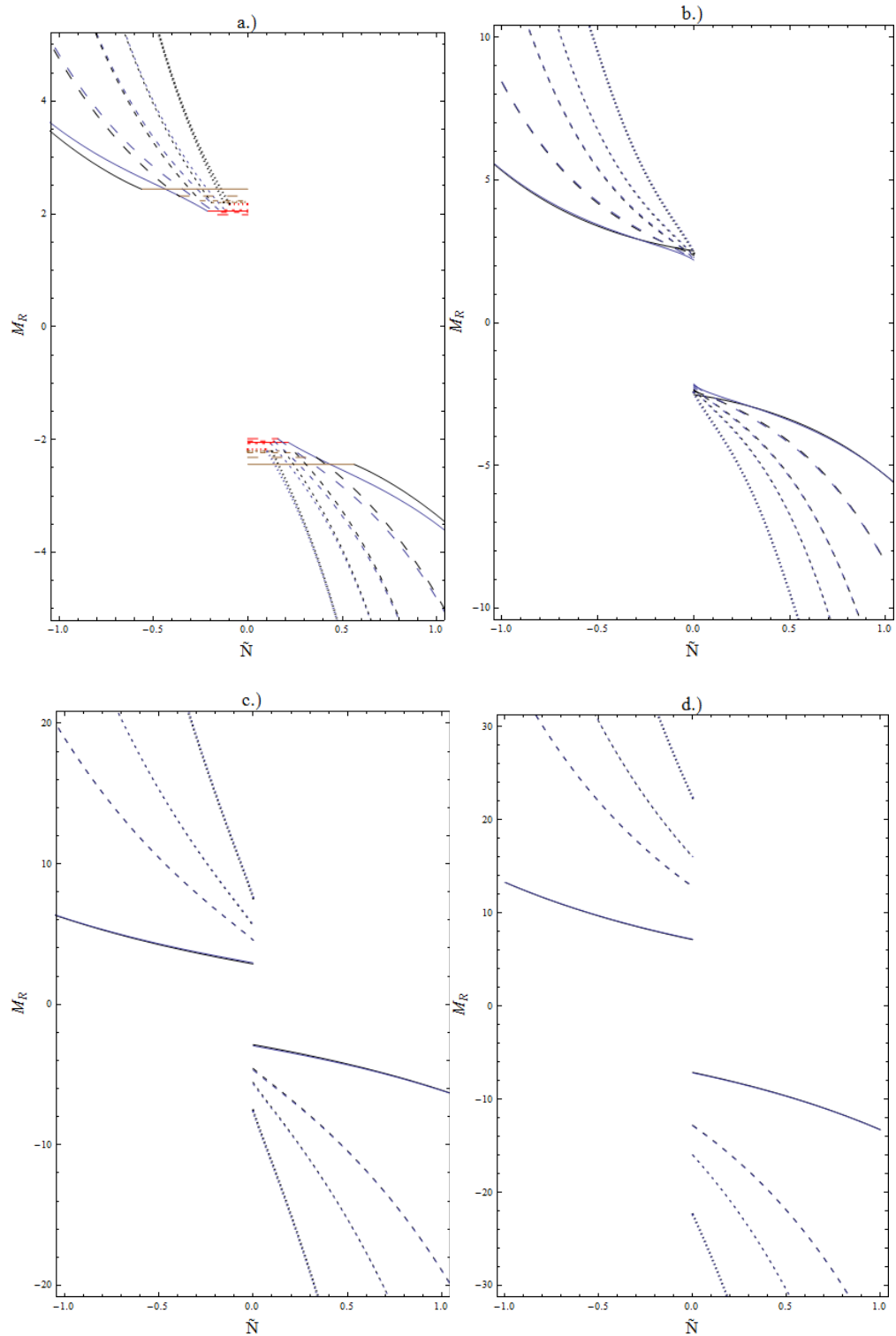

Fig.4. The moment $M$ (in units of $k_B T$), as a function of $\tilde{N}$, for various values of the ratio of the molecular steric radius $a$ to the bending persistence length $l_p$. We use the parameters in a.) $F_R = 25$ and $\beta = 0.1$, b.) $F_R = 25$ and $\beta = 0.35$, c.) $F_R = 500$ and $\beta = 0.1$, and d.) $F_R = 500$ and $\beta = 0.35$. Again, the black and brown (darker) curves correspond to the full model, whereas the red and blue (lighter) curves correspond to the simplifying approximation discussed at the end of Subsection 2.6 . The red and brown curves correspond to coexistence region between the braided state and unbraided state. The solid, long dashed, medium dashed, short dashed and dotted correspond to values $\tilde{a} = a/l_p = 1/50, 1/30, 1/20, 1/15, 1/10$, respectively. We see that the simplifying approximation works very well when $\beta = 0.35$ or $F_R = 500$, and always better for larger values of $\tilde{a}$.

In Fig. 4 we investigate the effect of changing the steric radius of the molecules on the applied moment as function of $\tilde{N}$. When we have coexistence, as seen in Fig. 4a for the values $\tilde{F}_R = 25$ and $\beta = 0.1$, the coexistence region decreases in size as we make $\tilde{a}$ larger, whereas $M_c$ changes little. By increasing $\tilde{a}$ (see Fig. 8 below) we push out the value of $\tilde{R}_0$ at fixed $M_R$, as we increase the amount of repulsion due to steric effects. Increasing $\tilde{R}_0$ reduces $\tilde{N}$, which is roughly proportional to $1/\tilde{R}_0$ (see [21]); therefore $\tilde{N}_c$ decreases. For the ground state this proportionality is exact with $\tilde{N} = \sin(\eta_0/2)/\pi\tilde{R}_0$. This also accounts for the increasing magnitude of $M_R$ at fixed $\tilde{N}$ with increasing $\tilde{a}$ in all the curves presented in Fig.4, outside the coexistence region.

### 3.2 Molecular Extension

We now examine how the molecular extension $\tilde{z}_T$, the ratio of the end to end distance to the total molecular contour length, changes with the number of braid turns. In Fig. 4, we fix $a/l_p = 1/50$ and plot the molecular extension for various values of the of the pulling force for both $\beta = 0.1$ and $\beta = 0.35$. As expected, as one increases the pulling force one increases the extension. It seems that the simplifying approximation always gives a lower value of the extension. Nevertheless, the simple approximation works well for $\beta = 0.1$ at large enough forces (the values $\tilde{F}_R = 100, 200, 400$) and for $\beta = 0.35$ for all the force values considered.

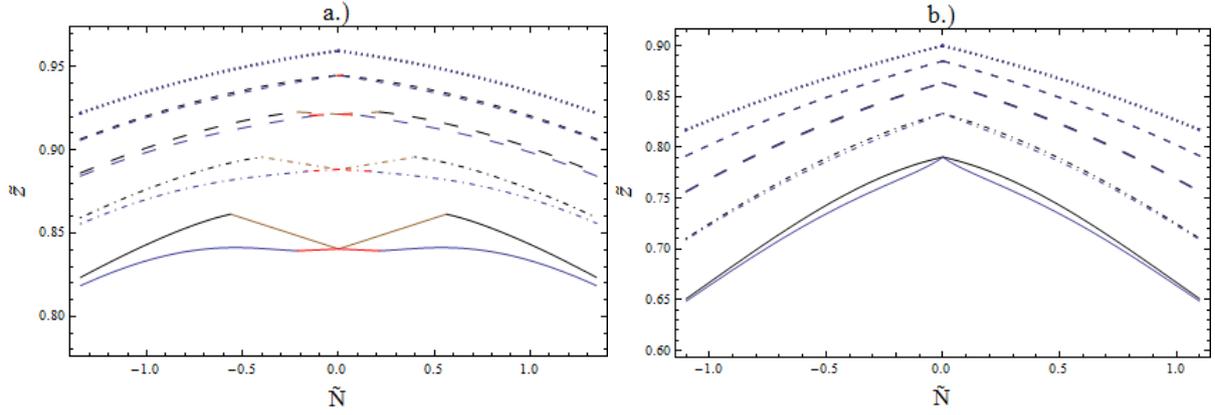

Fig. 5. The molecular extension $\tilde{z} = z/L$ as a function of number of braid turns per unit length. Plots are shown for a.) $\beta = b/L = 0.1$ and b.) $\beta = b/L = 0.35$. The colour coding is as in Fig. 3 and 4. The solid, dot dashed, long dashed, short dashed, and dotted lines correspond to rescaled pulling forces of $\tilde{F}_R = 25, 50, 100, 200, 400$.

In Fig. 4, we notice that the molecular extensions for given values of the pulling force are lower for $\beta = 0.35$ than for $\beta = 0.1$. Indeed, Eq. (1.34) shows that when the braid is just about to form, at $z_B = 0$, the extension goes down as $\beta$ is increased, arising simply from the geometrical considerations used to construct the model.

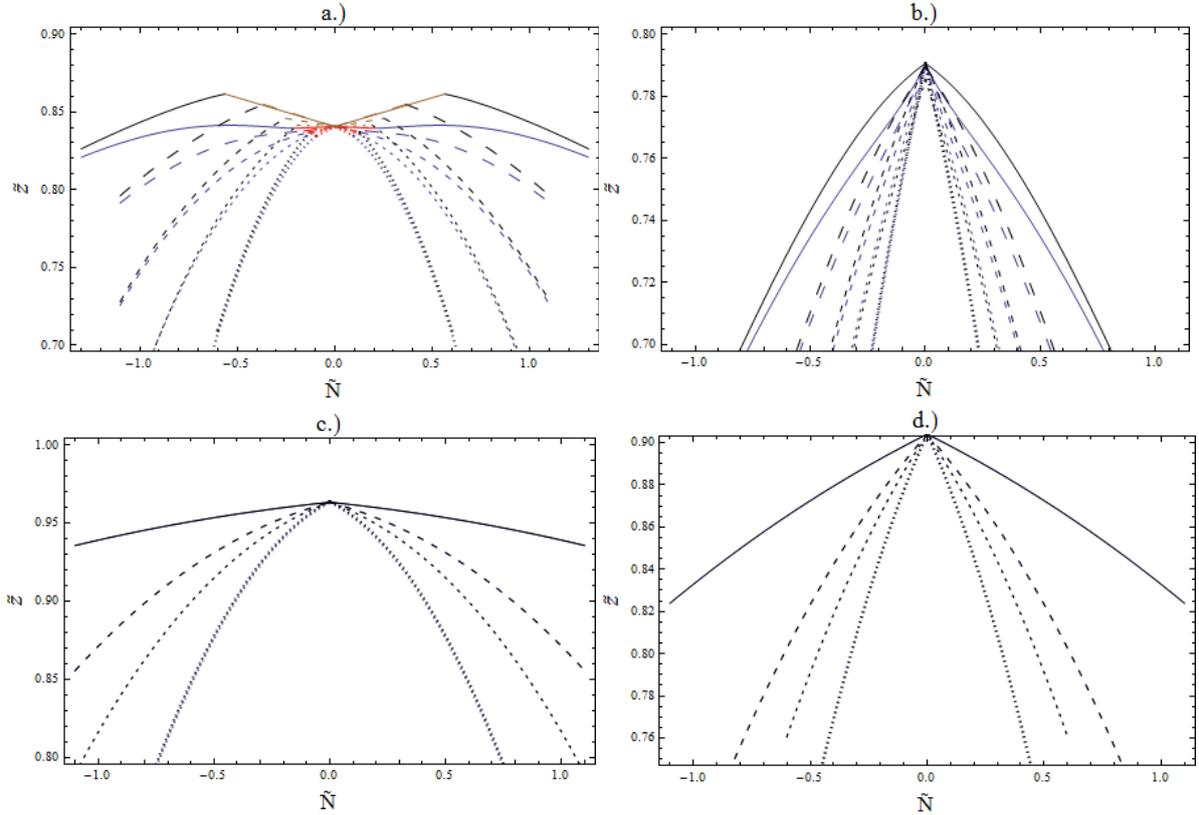

Fig.6. The molecular extension $\tilde{z} = z/L$ as a function of number of braid turns per unit length for various values of the steric radius $a$ to the persistence length $l_p$. The values of the other parameters used are: a.) $F_R = 25$, $\beta = 0.1$, b.) $F_R = 25$, $\beta = 0.35$ c.) $F_R = 500$, $\beta = 0.1$, and d.) $F_R = 500$, $\beta = 0.35$. The solid, long

dashed, medium dashed, short dashed and dotted correspond to values $\tilde{a} = a/l_p = 1/50, 1/30, 1/20,$ $1/15, 1/10$.

In Fig. 5 we examine how the molecular extension changes when we change $a$ the steric radius. As we increase the ratio $\tilde{a} = a/l_p$, the gradient of all the extension curves becomes steeper. This is a consequence of the braid radius becoming larger as $\tilde{a}$ is increased. We notice that as we increase $\tilde{a}$ the simple approximation works better at the low force value of $\tilde{F}_R = 25$, and always very well at $\tilde{F}_R = 500$.

### *3.3 Braid Geometric Parameters*

We now examine the geometric parameters of the braid, starting with the contour length $L_b$. In Fig. 7a, within the coexistence region (the red curves) $\gamma = L_b/L$ is constant, since the energy density and $\eta_{end}$ do not change. However, the proportions of $L_b$ that are braided and unbraided do. Note that, in all curves, $\gamma$ cannot exceed the value $1-\beta$, as $b/2$ is the minimum length that all four end pieces can take. We see that the gradient of $\gamma$ with respect to $\tilde{N}$ starts flattening, as we approach this limiting value. This is because $d\gamma/dM_R$ decreases as we increase $\gamma$, due to having to overcome the increasing entropic force that the end pieces exert, arising from their WLC fluctuations being suppressed as they yield contour length to the braid.

As we increase the pulling force, we see that $L_b$ is reduced at a fixed value of $\tilde{N}$. The pulling force tries to extend the end to end distance by reducing $\eta_{end}$, making the end pieces longer and tangents of their average centre lines more in line with the direction of the pulling force. Hence, to generate the same number of braid turns with a smaller value of $L_b$ the braid radius needs to become tighter as the pulling force is increased, which is what we indeed see in Fig.8.

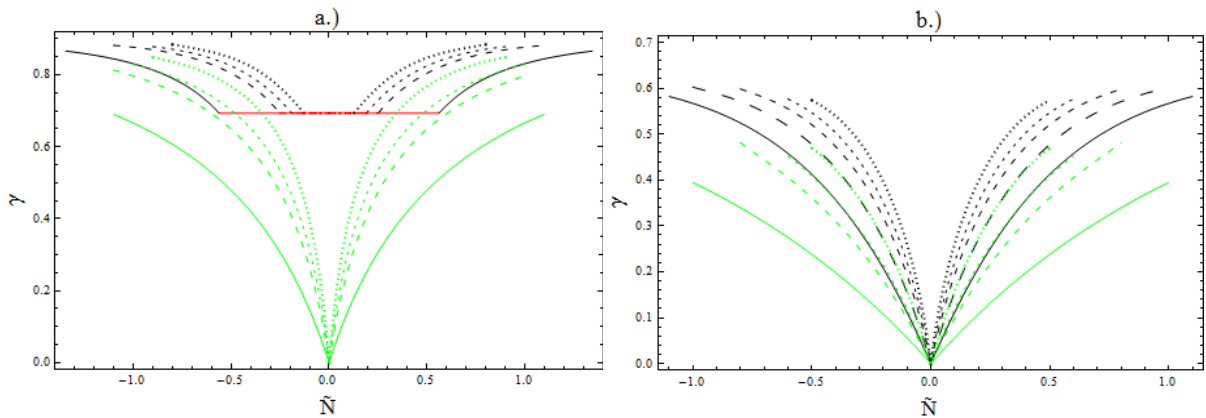

Fig.7. Plots showing the ratio of the contour length of the braided section to the total contour length of the molecules $\gamma = L_b/L$ as a function of the number of turns per persistence length $\tilde{N}$. Here, only the full model is used. In a.) the value $\beta = 0.1$ is used, the red and blue (dark) curves correspond to $\tilde{F}_R = 25$ and the green (light) curves to to $\tilde{F}_R = 500$. In b.) the value $\beta = 0.35$ is used, the blue (dark) curves correspond to $\tilde{F}_R = 25$

and the green (light) curves to to $\tilde{F}_R = 500$. The solid, long dashed, medium dashed, short dashed and dotted line correspond to $\tilde{a} = a/l_p = 1/50, 1/30, 1/20, 1/15, 1/10$, respectively.

In Fig. 8 we plot both the braid radius and tilt angle as functions of the number of braid turns per persistence length. In the coexistence region both $\eta_0$ and $R$ stay constant within the braided regions. However, in Fig. 8c, we plot the average value of $\bar{\eta}_0 = \eta_0 \tilde{N} / |\tilde{N}_c|$. As we increase the number of turns outside the coexistence region, the braid radius decreases and $\eta_0$ increases. This is because $L_b$ does not increase in a linear fashion, as discussed above. Therefore, to accommodate an increasing number of braid turns, both of these quantities must change in this way.

When we examine $\tilde{R}_0$, we see that for low forces it changes considerably, whereas for larger forces it stays roughly constant. This can be explained by the fact that as we increase $M_R$, to increase the number of braid turns the braid must tighten. However, steric forces resist this tightening. When the molecules come close enough that $\tilde{R}_0 \sim \tilde{a}$, $d\tilde{R}_0 / dM_R$ becomes very small; steric forces increase rapidly as $\tilde{R}_0$ approaches $\tilde{a}$. At large forces, as we already discussed, to stabilize the braid we require a large value of $M_R$. A larger value $M_R$ forces the molecules in the braid closer together. Then, as we increase $M_R$ above its large threshold ($\tilde{N} = 0$) value the radius of the braid can change little. The increasing steric forces, with increasing $\tilde{a}$, push the two molecules apart, so that $\tilde{R}_0$ increases as a function of $\tilde{a}$. The steric forces also ensure we can never have $\tilde{R}_0 \leq \tilde{a}$. We see that increasing $\beta$, at large forces, makes $\tilde{R}_0$ change even less with respect to $\tilde{N}$. This is because the moment required to generate a braid is even larger.

As we increase the force in Fig. 8c we see that the tilt angle decreases slightly. This favours a larger value of $\tilde{z}_B$, contributing to an increasing molecular extension $\tilde{z}_T$ with increasing pulling force. However, in Fig. 8d we see that this trend is reversed. This feature may be attributed to a competing effect. If $L_b$ decreases significantly as the pulling force is increased, at fixed $\tilde{N}$, to conserve the number of braid turns, as well as reducing $\tilde{R}_0$ one may also need to increase the tilt angle $\eta_0$. Indeed, if we look at Fig. 7, we see that at $\beta = 0.35$ the value of $\gamma$ for the two force values is smaller than it is for $\beta = 0.1$.

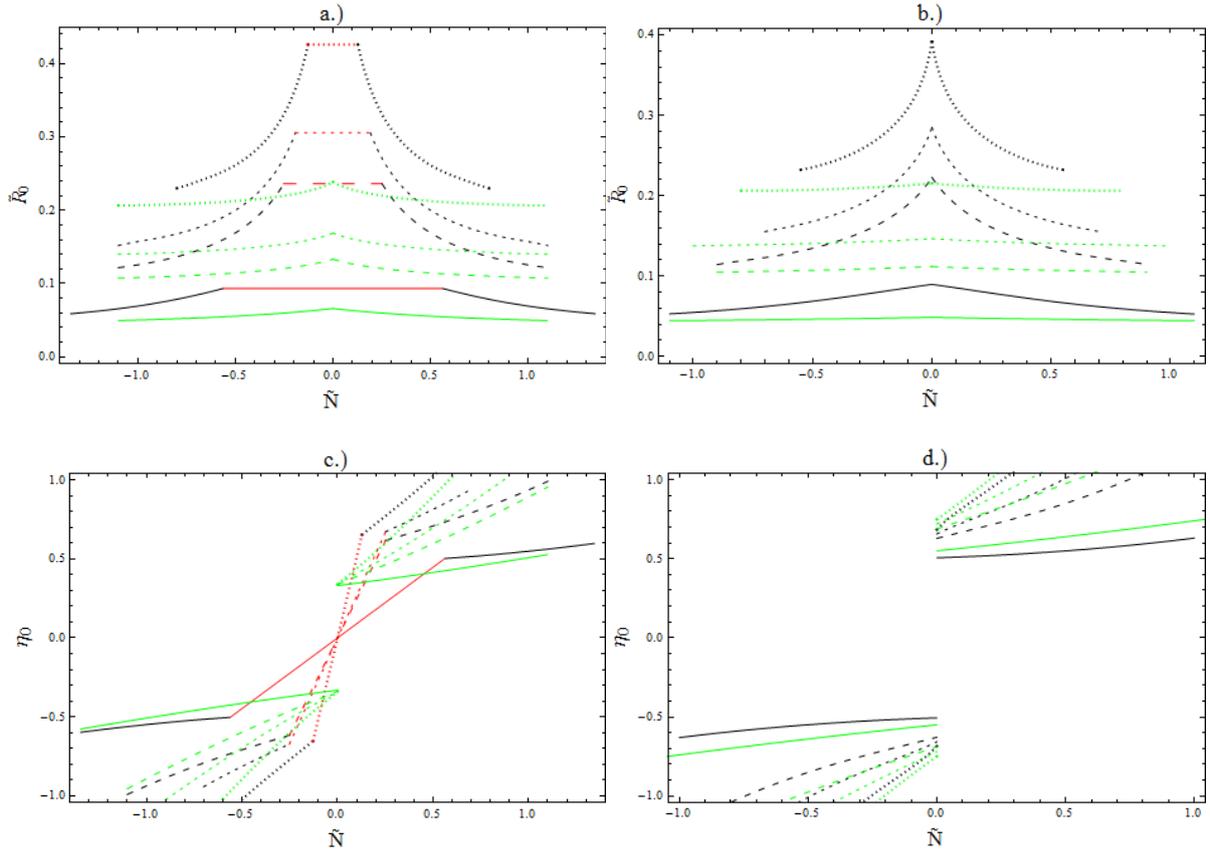

Fig,8. Figures showing both the rescaled braid radius $\tilde{R} = R/l_p$ and tilt angle $\eta_0$ as functions of the number of turns per persistence length. Here only the full model is used. Shown in a.) and b.) is the braid radius and in c.) and d.) the tilt angle. In a.) and c.) the value $\beta = 0.1$ is used. In b.) and d.) the value $\beta = 0.35$ is used. In all plots red and black (dark) curves correspond to $F_R = 25$ and the green (light) curves to to $F_R = 500$. The solid, long dashed, medium dashed, short dashed and dotted line correspond to $\tilde{a} = a/l_p = 1/50, 1/30, 1/20, 1/15, 1/10$, respectively.

### 3.4 Lateral force holding ends apart

Last of all, we examine how the rescaled lateral force that holds the two ends apart $F_b$ changes with the number of braid turns. Firstly, looking at Fig.8a it seems that for $\tilde{F}_R = 25$, the lateral force, when we have coexistence, is independent of $\tilde{a}$ at a value of $F_b \simeq 10$. This independence with respect to the steric radius can be understood in the following way. The unbraided state free energy depends only on $F_R$ (see Eq. (1.36)), whereas the free energy of the braided state depends also on $M_R$ and $\tilde{a}$. However, at coexistence, Eq. (1.37) is satisfied, and so at $M_R = M_c$ the energy density does actually not depend on $\tilde{a}$. Now, if we examine Eq. (1.31) we see that $\eta_{end}$ depends on the ratio of the pulling force to the energy density. Therefore, if the energy density in the coexistence region depends only on $F_R$, through Eq. (1.4) the lateral force $F_b$ depends only on $F_R$.

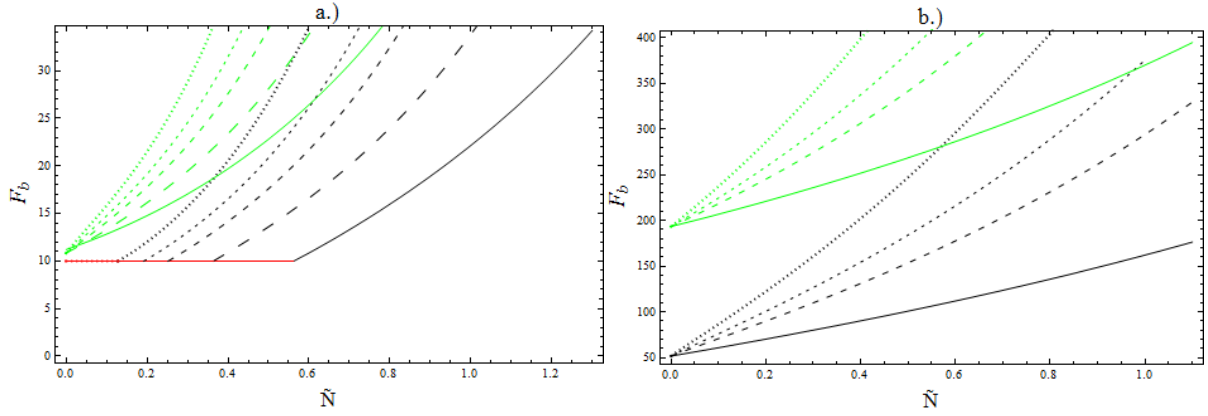

Fig. 9. Figures showing the lateral force needed to keep both ends at distance $b$ apart as a function of the number of braid turns per persistence length, $\tilde{N}$. In a.) the rescaled force value $\tilde{F}_R = 25$ is used, and in b.) the value $\tilde{F}_R = 500$. The green (light) curves correspond to $\beta = 0.35$ and the black and red (dark) curves correspond to $\beta = 0.1$. The solid, long dashed, medium dashed, short dashed and dotted line correspond to $\tilde{a} = a/l_p = 1/50, 1/30, 1/20, 1/15, 1/10$, respectively.

We see that increasing the pulling force and increasing $\beta$ both increase $F_b$ at a fixed number of braid turns. The latter can be explained by examining the point where a braid is just about to form, at $L_b = 0$. The value of $\eta_{end}$ at this point, $\eta_{min}$ increases as $\beta = b/L$ is increased. We see from simple geometric considerations, through Eq. (1.4), that if we keep $F_R$ fixed and increase $\eta_{end}$ $F_b$ must increase. In general, for a fixed number of turns, increasing $\beta$ always increases $\eta_{end}$. As we make $\tilde{a}$ larger we increase the size of $F_b$ at fixed $\tilde{N}$. Firstly, the size of $\eta_{end}$ mainly depends on both $M_R$ and $F_R$. Secondly, since $\tilde{R}_0$ increases with increasing $\tilde{a}$, for a fixed value of $M_R$, $\tilde{N}$ decreases. Thus, at fixed $\eta_{end}$, $\tilde{N}$ decreases, so accounting for this effect.

## 3. Discussion and outlook

In this paper we have developed a model describing the braiding of two semi-flexible rod like molecules with negligible finite ranged interactions between them. The model, based on a statistical theory of braiding, makes predictions for braiding experiments similar to those conducted for DNA in Refs. [18,19]. Testing these predictions, for the braiding of such molecules, might be useful in isolating and understanding the important factors influencing braiding in biological and manmade materials.

We see that the simplifying approximation, where the average values of the bending energy and work terms are replaced with their values at the average braid configuration, works well provided that either the distance between the two molecular ends $b$, or the pulling force, is large enough. This approximation was utilized in Ref. [20], at $b \approx 0.35L$ ($L$ is the contour length of the DNA), therefore it seems that this approximation is justifiable here, when the molecules are sufficiently apart for interactions between them to be considered weak. However, it was pointed out in Ref. [20], that a self-consistent determination of the mean squared amplitude (discussed below) might fit experimental data better at larger values of $|N|$, where the molecules come closer together.

One thing that is missing from the whole approach is braid buckling. We expect that at a fixed value of $F_R$, as we increase $M_R$, we would expect a buckling transition to a state where the average braid axis is no-longer straight. As we increase $\tilde{N}$, we observe that $\eta_0$ and $\eta_{end}$ increase. For two braided rods in steric contact $\eta_0$ and $\eta_{end}$ cannot exceed the value of $\pi/2$. Thus, when this value is reached, to increase $Lk_b$ (or $\tilde{N}$), the braid must buckle. With thermal fluctuations the two molecules lie apart, so that $\tilde{R}_0 > \tilde{a}$. Therefore, we can allow for $\eta, \eta_{end} > \pi/2$ (though some of the initial considerations and the energy functional in terms of the geometric parameters for the elastic energy shown in Ref. [21] are not strictly valid). However, we would expect that steric forces are very much increased in this case making this situation less favourable than the extra bending energy needed for buckling. Furthermore, as we increase $\tilde{N}$ we decrease $\tilde{R}_0$, thereby again increasing the free energy due to steric interactions making the buckled state more favourable. On buckling, one thing we might expect is the formation of a super plectoneme like structure(s), where the braid centre line wraps around its self, tracing out a plectoneme. Modelling such structures has already been considered in Ref. [14], and the approach used there could be adapted to extend our model to consider such buckling. It would also be useful to compare this super plectoneme state with a state where the braid axis forms a solenoid.

The model can also be simply modified to include a weak interaction potential between the molecules by simply adding an interaction energy term that may depend on the average 'braid radius' $R_0$ and tilt angle $\eta_0$. In other words, we can add an interaction term $V(R_0, \eta_0)$ to the energy density described by Eq. (1.26). This was precisely what was done in Ref. [20], when considering DNA molecules under conditions where chiral forces depending on the DNA helical structure are weak. On the other hand, if the interactions are sufficiently strong, $V(R_0, \eta_0)$ starts to play a role in determining the extent of the fluctuations in $R(\tau)$; more precisely the value of $\langle (R(\tau) - R_0)^2 \rangle$. In this case, to include interactions, $\langle (R(\tau) - R_0)^2 \rangle$ must be determined self-consistently by $V(R_0, \eta_0)$ as well as the steric interaction. Already, in Ref. [21], we have built an variational approximation that does this, using an approach similar to Ref. [13], which we hope to exploit in later work.

Another possible extension to this work would be to consider the molecules as no longer being smooth cylinders. In Ref. [16] it was discussed how one could take account of helical molecular shape in the steric interaction. This could be done by making $\alpha_H$ depend on the average azimuthal orientation of the molecules (see Ref. [16]). However, such an extension has yet to be achieved.

In two subsequent papers we will give results for two different cases with inter-molecular interactions. In the first paper, the interactions will be purely repulsive and electrostatic, and we will fit our results to the existing experimental data of Ref. [19], to improve fits of Ref. [20]. In the second paper, we will deal with interactions with an attractive component in the interaction. This additional term leads to the possibility of a collapse of the braid into a tighter structure.

## Acknowledgements

D.J. Lee would like to acknowledge useful discussions with R. Cortini , A. Korte, A. A. Korynshev, E. L. Starostin and G.H.M. van der Heijden. This work was initially inspired by joint work that has been supported by the United Kingdom Engineering and Physical Sciences Research Council (grant EP/H004319/1). He would also like to acknowledge the support of the Human Frontiers Science Program (grant RGP0049/2010-C102).